\documentclass[a4paper,11pt]{article}

\usepackage[a4paper,left=3cm,right=3cm,top=3cm,bottom=2.5cm]{geometry}
\usepackage{graphicx} 
\usepackage{authblk}
\usepackage{color} 
\usepackage{amsmath}
\usepackage{amssymb} 
\usepackage{hyperref} 
\hypersetup{
	colorlinks=true,
	citecolor=black,
	filecolor=black,
	linkcolor=blue,
	urlcolor=blue
}
\usepackage[labelfont=bf]{caption}

\title{\texttt{W0\_sample = np.random.normal(0,1)?}}
\author[a]{J. Ebelt}
\author[a,b]{S. Krippendorf}
\author[c]{A. Schachner}

\affil[a]{\footnotesize Arnold Sommerfeld Center for Theoretical Physics, Ludwig-Maximilians Universität, \protect\\ Theresienstr.~37, 80333 München, Germany}
\affil[b]{\footnotesize Universitäts-Sternwarte, Fakultät für Physik, Ludwig-Maximilians Universität,\protect\\
Scheinerstr.~1, 81679 München, Germany}
\affil[c]{\footnotesize Department of Physics, Cornell University, Ithaca, NY 14853, USA}
\date{\today}

\begin{document}
\begin{flushright}
LMU-ASC 27/23
\end{flushright}
{\let\newpage\relax\maketitle}

\begin{abstract}
    In this note we explore the distribution of vacuum expectation values of the superpotential $W_0$ in explicit Type~IIB flux compactifications. We show that the distribution can be approximated, universally across geometries, by a two-dimensional Gaussian with a model dependent standard deviation. We identify this behaviour in $20$ Calabi-Yau orientifold compactifications with between two and five complex structure moduli by constructing a total of $\mathcal{O}(10^7)$ flux vacua. We observe a characteristic scaling behaviour of the width $\sigma$ of our distributions with respect to the D3-charge contributions $N_{\text{flux}}$ from fluxes which can be approximated by $\sigma \sim \sqrt{N_{\text{flux}}}$. This $W_0$ distribution implies that locating small values of $|W_0|$ as a preferred regime associated with classes of string theory solutions typically featuring hierarchies, simplifies to the basic statement of finding small Euclidean norms of normally distributed values. We do also identify small modifications to this Gaussian behaviour in our samples which might be seen as indications for the breakdown of the continuous flux approximation commonly used in the context of statistical analyses of the flux landscape.
\end{abstract}

\section{Introduction}

The distributions of observables arising in effective field theories (EFTs) from string theory hold vital information about the attainable low-energy physics within string theory. This includes insights into the genericity of low-energy phenomena, as well as into the distinguishing features of string theory constructions from standard bottom-up EFT models. On a higher level, such distributions can be linked with vacuum selection mechanisms of vacua resembling the physics we see around us.

Flux compactifications of Type~IIB string theory provide the perfect arena to study the distribution of such low-energy properties. Here, one investigates the four-dimensional scalar potential for complex structure moduli and the axio-dilaton induced by 3-form fluxes~\cite{Giddings:2001yu} (see~\cite{Grana:2005jc,Douglas:2006es,Hebecker:2021egx} for reviews). In particular, we are interested in understanding minima satisfying $F$-flatness conditions $D_I W=0$ in the large complex structure regime where moduli potentials are under computational control \cite{Hosono:1994av,Hosono:1994ax}. Recent numerical improvements~\cite{Dubey:2023dvu} enable us to get sizable samples of such solutions across a variety of geometries at relatively small computational costs. In this setting, we look at the distribution of\footnote{Here, $K$ denotes the Kähler potential and $W_{\text{flux}}$ the flux induced superpotential \cite{Gukov:1999ya}.}
\begin{equation}\label{eq:w-complex}
    W_0=\sqrt{\frac{2}{\pi}}\left\langle e^{K/2} W_{\text{flux}}\right\rangle~,
\end{equation}
which is tightly related to the gravitino mass.

\begin{table}[t!]
    \centering
    \begin{tabular}{c|c|c|c}
         $h^{1,1}$ & $h^{1,2}$ & $Q_{D3}$ & $\sharp$vacua  \\
         \hline
         \hline
          144& 2 & 148  & 869,503 \\
         \hline
          120 & 2 & 124  & 776,494 \\
         \hline
          132 & 2 & 136  & 842,102 \\
         \hline
          128 & 2 & 132 & 899,324\\
         \hline
          272 & 2 & 276 & 1,106,802\\
    \end{tabular}
    \hspace{0.5cm}
    \begin{tabular}{c|c|c|c}
         $h^{1,1}$ & $h^{1,2}$ & $Q_{D3}$ & $\sharp$vacua  \\
         \hline
         \hline
          99 & 3 & 104  & 707,365 \\
         \hline
          115 & 3 & 120  & 1,518,797 \\
         \hline
          107 & 3 & 112 & 838,071 \\
         \hline
          119 & 3 & 124 & 996,795\\
         \hline
          243 & 3 & 248 & 1,666,153\\
    \end{tabular}
    
    \vspace{0.5cm}
    
    \begin{tabular}{c|c|c|c}
         $h^{1,1}$ & $h^{1,2}$ & $Q_{D3}$ & $\sharp$vacua  \\
         \hline
         \hline
          122 & 4 & 128 & 1,222,120 \\
         \hline
          130 & 4 & 136 & 258,795 \\
         \hline
          118 & 4 & 124 & 181,666 \\
         \hline
          142 & 4 & 148 & 520,795 \\
         \hline
          98 & 4 & 104 & 943,414\\
    \end{tabular}
    \hspace{0.5cm}
    \begin{tabular}{c|c|c|c}
         $h^{1,1}$ & $h^{1,2}$ & $Q_{D3}$ & $\sharp$vacua  \\
         \hline
         \hline
          149 & 5 & 156 & 209,870 \\
         \hline
          153 & 5 & 160 & 228,429 \\
         \hline
          165 & 5 & 172 & 310,252 \\
         \hline
          89& 5 & 96 & 1,242,731 \\
         \hline
          213 & 5 & 220 & 146,970 \\
    \end{tabular}
    \caption{Summary of our compactification manifolds with their respective Hodge numbers, tadpole values and number of vacua. In total, we find $15,486,448$ solutions.}
    \label{tab:models}
\end{table}

A priori, it is unclear how $W_0$ is distributed as it is only determined upon solving a system of polynomial equations with an infinite sum of exponential corrections. In the case of large complex structure limits, the coefficients of these equations are given by topological quantities arising from the respective mirror geometries~\cite{Hosono:1994av,Hosono:1994ax}, see also \cite{Demirtas:2023als} for recent advances. Ultimately, we would like to address the question of whether these types of structures of string theory leave imprints on the distributions of phenomenological observables.

In this context, it has been argued that many distributions depend only on a ``few'' UV parameters \cite{Douglas:2003um,Ashok:2003gk,Denef:2004ze,Denef:2008wq} like the orientifold tadpole as will be partially confirmed within this note. In particular, it has been previously argued that $W_0$ is distributed as a Gaussian with a tadpole dependent width. It is important to stress though that the aforementioned analyses have been mostly performed for hypothetical geometries in the continuous flux approximation. We will instead use geometries with explicit $\mathbb{Z}_2$-involutions for the orientifold and construct actual vacua obtained from solving $F$-term conditions for quantised fluxes. This allows us to obtain actual sizable samples of flux vacua from first principles, enabling actual statistical analysis of {\it string data}.

In this quest, we find that -- as a first approximation -- the real and imaginary part of $W_0$ indeed appear to be distributed as a Gaussian and that deviations arise only at sub-leading order. Our evidence is based on $20$ background geometries for which we generated reasonably sized samples (see Table~\ref{tab:models} for details). In this dataset, we find universal behaviour across the various geometries. From the physics perspective, this means that the $W_0$-distributions arising from string theory, at least in our current samples, are indeed simple and almost universal. Furthermore, these observations allow for projections on the required sample size needed to generate solutions with small absolute values for $|W_0|$ which is an essential ingredient for the hierarchical scale separation in the KKLT scenario~\cite{Kachru:2003aw}. Assuming that $W_0$ is normally distributed, the respective standard deviation of this distribution sets the expected hierarchical suppression.

The rest of this note is organised as follows: in Section~\ref{sec:geometries} we describe our sample of geometries and algorithmic choices for fluxes. We present our numerical results in Section~\ref{sec:numericalresults} and conclude in Section~\ref{sec:conclusions}.

\section{Geometry and flux sample}
\label{sec:geometries}

\begin{figure}[t!]
\begin{center}\includegraphics[width=1\textwidth]{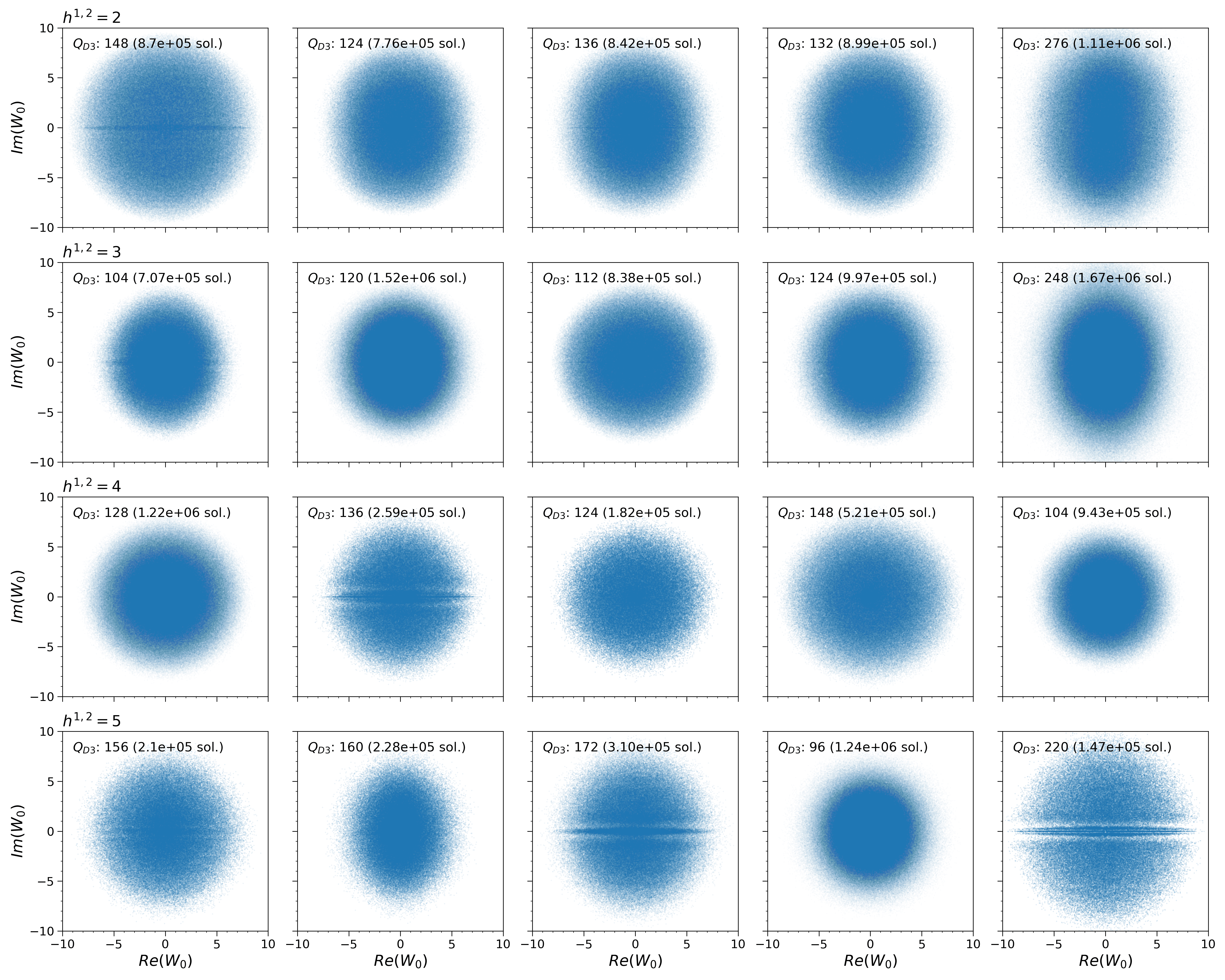}
\end{center}
\caption{Distribution of our solutions for the real and imaginary parts of $W_0$. Each box corresponds to one of our geometries as listed in Tab.~\ref{tab:models}.}\label{fig:gaussian_ensemble}
\end{figure}

To produce sufficiently large samples of flux vacua, we pick 20 CY orientifold models with $h^{1,2}=2,3,4,5$ as summarised in Tab.~\ref{tab:models} for each of which we construct at least $\mathcal{O}(10^5)$ vacua. We compute the tadpole values $Q_{D3}=h^{1,1}+h^{1,2}+2$ from the corresponding orientifold configurations with $h^{1,2}_+=0$ following the algorithm of~\cite{Moritz:2023jdb}. The D7-tadpole is cancelled locally by putting D7-branes on top of the O7-planes.\footnote{Details about the geometries are included in the ancillary files associated to this note.} Using \texttt{CYTools}~\cite{Demirtas:2022hqf,Demirtas:2023als} we calculated the prepotential including the instanton series up to degree $10$. We follow the conventions described in~\cite{Dubey:2023dvu} to which we refer for more details.

We search for solutions to the $F$-term conditions $D_I W=0$ for all complex structure moduli $Z^i$, $i=1,\ldots,h^{1,2}$, and the axio-dilaton $\tau$.
To solve this optimisation problem, we generate fluxes and starting points for the minimisation using our ISD$_+$ sampling technique described in \cite{Dubey:2023dvu} by restricting to choices with $N_{\text{flux}}\leq Q_{D3}$.
Specifically, we sample half of the flux vector from the range $[-5,5]$ and our starting points for the moduli within the (mirror) Kähler cone subject to the cutoff conditions
\begin{equation}
    |\text{Im}(Z^i_0)|,\,|\text{Im}(\tau_0)|\leq 5\; , \quad |\text{Re}(Z^i_0)|,\,|\text{Re}(\tau_0)|\leq 0.5\, .
    \label{eq:samplingrange}
\end{equation}

On the resulting vacuum expectation values (VEVs) for the moduli, we subsequently evaluate EFT quantities of interest such as the flux superpotential $\langle W\rangle$. The latter by itself is however not necessarily meaningful due to Kähler transformations. To avoid this we re-scale the superpotential with the Kähler potential and define $W_0$ as in Eq.~\eqref{eq:w-complex} (cf.~\cite{Denef:2004ze}). Below, we will report the $W_0$ distribution and examine its general behaviour across geometries. 
Subsequently, we focus on the gauge invariant quantity of the absolute value of $|W_0|$. We leave a discussion of the physics of the phase of $W_0$ for the future (see for instance~\cite{Endo:2005uy} for a discussion of phases in the context of soft-supersymmetry terms). 

\section{Numerical results}
\label{sec:numericalresults}

\begin{figure}[t!]
\begin{center}
\includegraphics[width=1\textwidth]{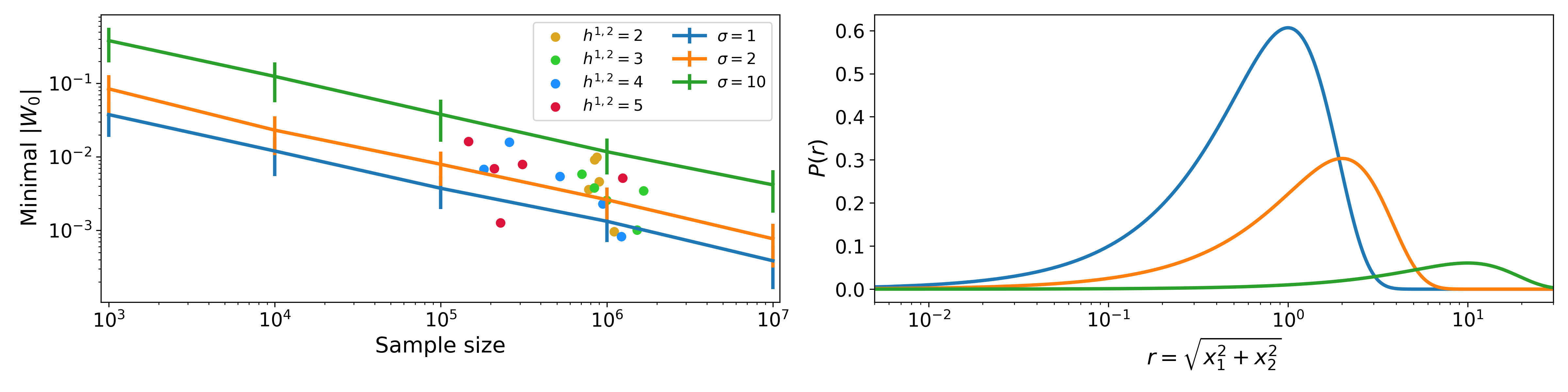}
\includegraphics[width=1\textwidth]{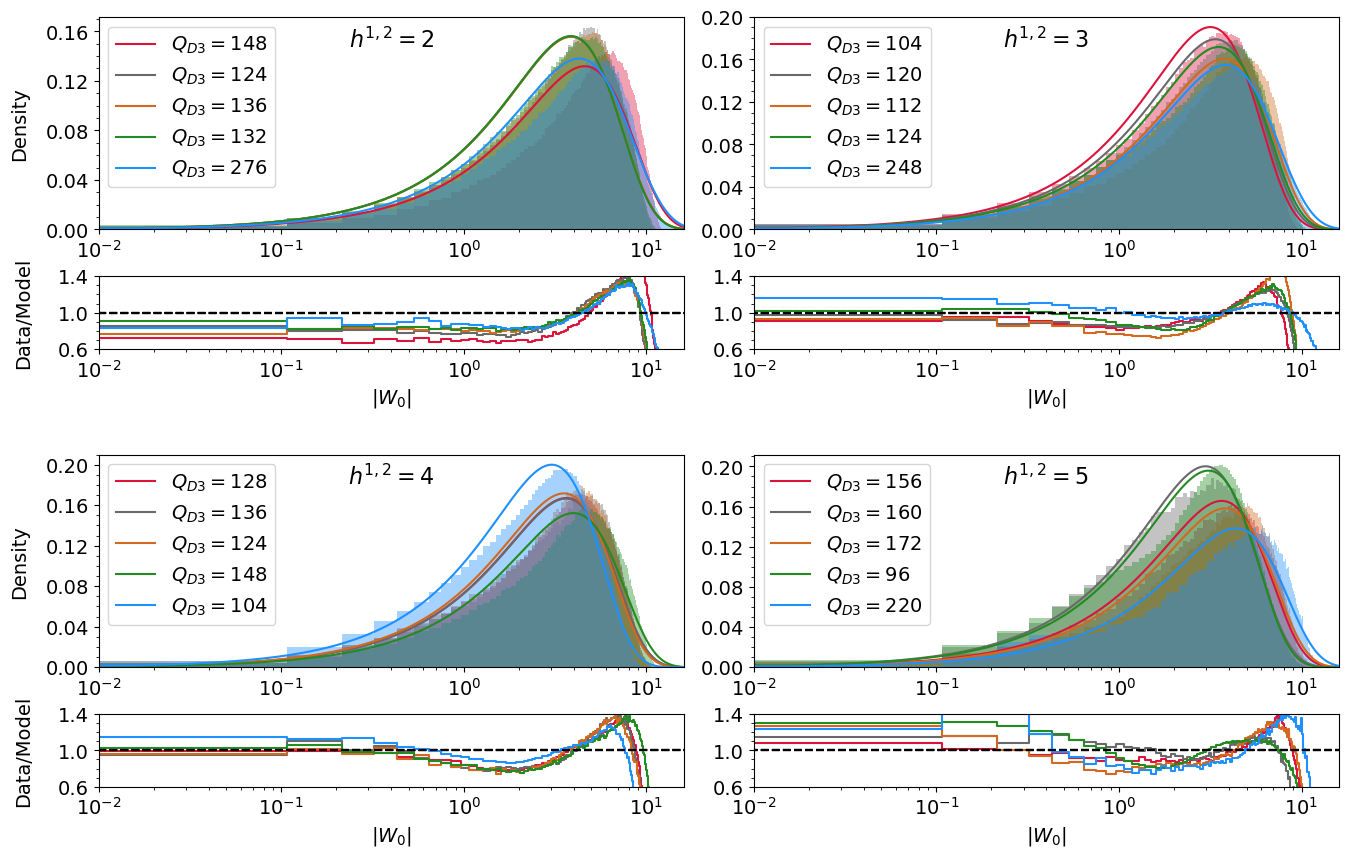}
\end{center}
\caption{\emph{Top left}: Comparison of expected minimal $|W_0|$ value as a function of the sample size. Different lines correspond to two-dimensional Gaussians with different standard deviations. The error bars are one standard deviation and are estimated with $100$ runs each. The dots indicate our 20 samples coloured by their respective values of $h^{1,2}$. \emph{Top right:} The distribution of the absolute value for a two-dimensional Gaussian for different values of the standard deviation.  
\emph{Middle and bottom row}: The distributions of $|W_0|$ in our flux vacua samples exhibit universal behaviour across geometries. We show the best fit Gaussian associated to our data sample and the ratio of this model with respect to our data.
}\label{fig:comparison}
\end{figure}

We now turn to a discussion of our empirical findings. We show the respective distributions of the re-scaled superpotential in Figure~\ref{fig:comparison}. As previously advertised, we find a universal behaviour across the different geometries with varying number of moduli. The distribution looks remarkably similar to that of a two-dimensional Gaussian which we discuss shortly. It is also astonishing that all of these solutions fall in a comparable range of values, indicating rather similar standard deviations for the Gaussians.

We note that some of the geometries show a feature close to ${\rm Im}(W)=0$ which corresponds to values where a large re-scaling with the Kähler potential occurs. After further inspection, we discovered no inconsistencies for these solutions which is why we attribute these structures to sampling and algorithmic biases at this stage.

\begin{figure}[t!]
\begin{center}
\includegraphics[width=1\textwidth]{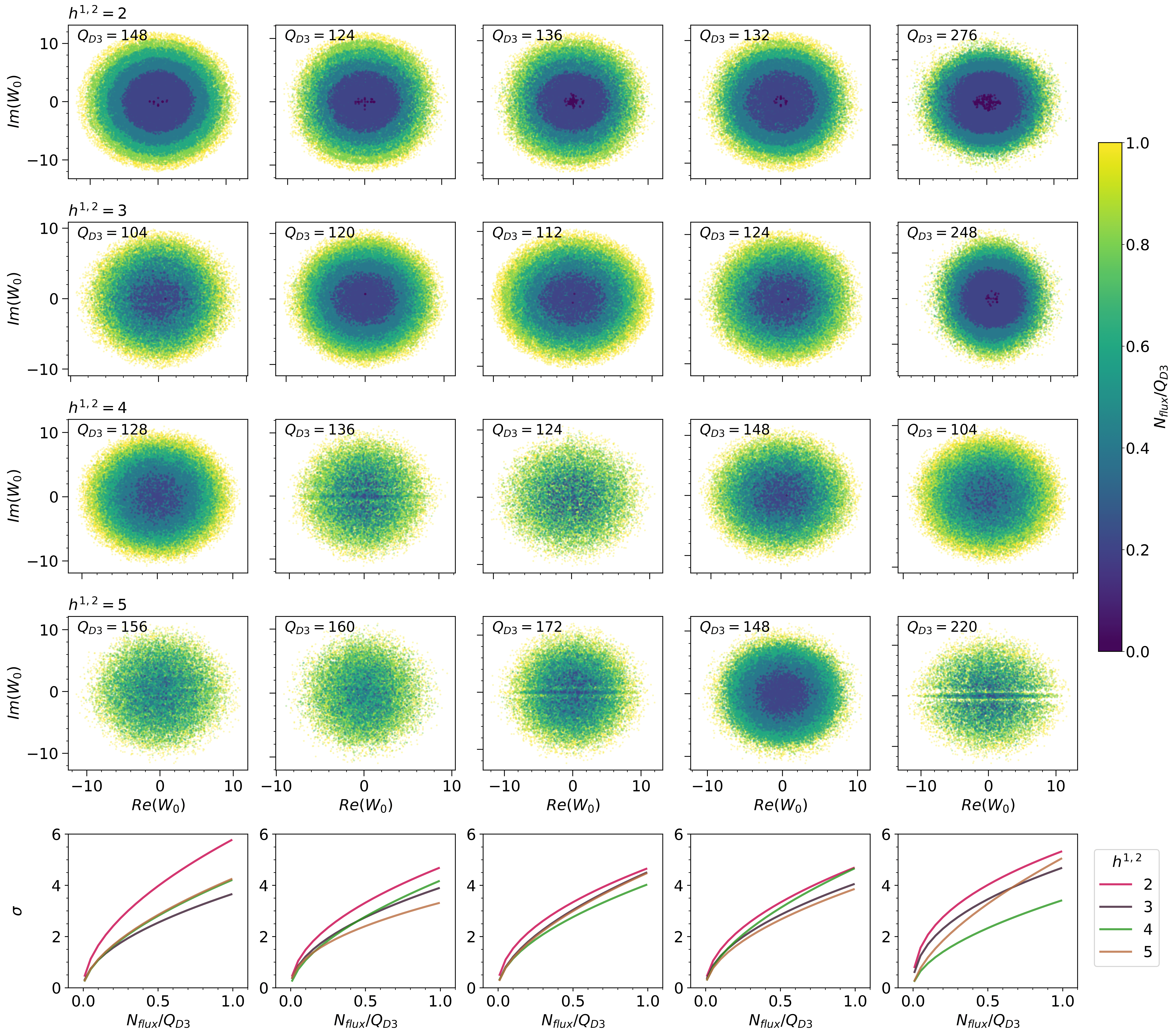}
\end{center}
\caption{Scaling of the distribution of $W_0$ on $N_{\text{flux}}$. Each box corresponds again to one of our geometries as listed in Tab.~\ref{tab:models}. The colors in the first four rows highlight the dependence on the D3-charge contribution  $N_{\text{flux}}$ from fluxes. For clarity, we picked only discrete values for $N_{\text{flux}}/Q_{D3}\in [0.01,0.2,0.4,0.6,0.9,1]$ to show the qualitative dependence of the width on $N_{\text{flux}}$. The bottom row shows the standard deviation $\sigma$ as a function of $N_{\text{flux}}/Q_{D3}$.}\label{fig:gaussian_ensemble_colored}
\end{figure}

The approximately Gaussian behaviour for $W_0$ was predicted by the findings of \cite{Denef:2008wq} where it was argued that, at least under simplifying assumptions including the continuous flux approximation, $W_0$ should be distributed as a Gaussian distribution peaked around zero with the standard deviation $\sigma$ being proportional to the tadpole $Q_{D3}$.

In fact, we observe such a scaling of the width in our distributions, i.e.,~we find that it scales with $N_{\text{flux}}/Q_{D3}$. This is shown in Fig.~\ref{fig:gaussian_ensemble_colored} where the colours highlight samples with different ratios $N_{\text{flux}}/Q_{D3}$. To make this quantitative, the last row in Fig.~\ref{fig:gaussian_ensemble_colored} shows $\sigma$ as a function of $N_{\text{flux}}/Q_{D3}$ for the various models. We clearly see a universal trend for the slope across the different examples. Empirically, we find that
\begin{equation}
    \sigma\sim \sqrt{\dfrac{N_{\text{flux}}}{Q_{D3}}}
\end{equation}
is an excellent approximation for most of our models in agreement with \cite{Denef:2008wq}.

Next, let us focus on the absolute value $|W_0|$ which is related to the value of the gravitino mass. Assuming a Gaussian distribution for the complex value, the resulting distribution for the absolute value is shown in Fig.~\ref{fig:comparison} in the top right. At an elementary level, the observed minimal value of $|W_0|$ scales inversely with the sample size. For illustrative purposes we show the respective dependence for various standard deviations. The minimal values we identify in our samples are highlighted in Fig.~\ref{fig:comparison} and they deviate only slightly from the Gaussian expectation. We also observe no noticeable trend with the number of moduli or the tadpole in these minimal values at this level. We hence expect that simply generating more samples will lead to the generation of solutions with larger hierarchical suppression. Having said that, an order of magnitude suppression requires roughly a two orders of magnitude increase in the sample size. Hence without using additional search strategies on the UV side (see~\cite{Demirtas:2019sip,Demirtas:2020ffz,Alvarez-Garcia:2020pxd} for human strategies), the expected cost for finding flux samples with a particular hierarchical suppression increases exponentially with the sample size. Similar to finding solutions with small tadpole at $h^{1,2}\gg 1$, we strongly believe that more time needs to be devoted to developing algorithmic approaches to probe exponentially suppressed tails of distributions within the landscape.

Moving to a quantitative comparison, we show in the bottom two rows of Figure~\ref{fig:comparison} the best-fit Gaussian distribution and the respective ratio between our data and this model. We make out two features, namely the fits almost consistently under-predict at smaller values in comparison to our data, while they over-predict at larger values. These deviations from a simple Gaussian will be analysed in more detail in future works.

\begin{figure}[t!]
    \centering
    \includegraphics[width=1\textwidth]{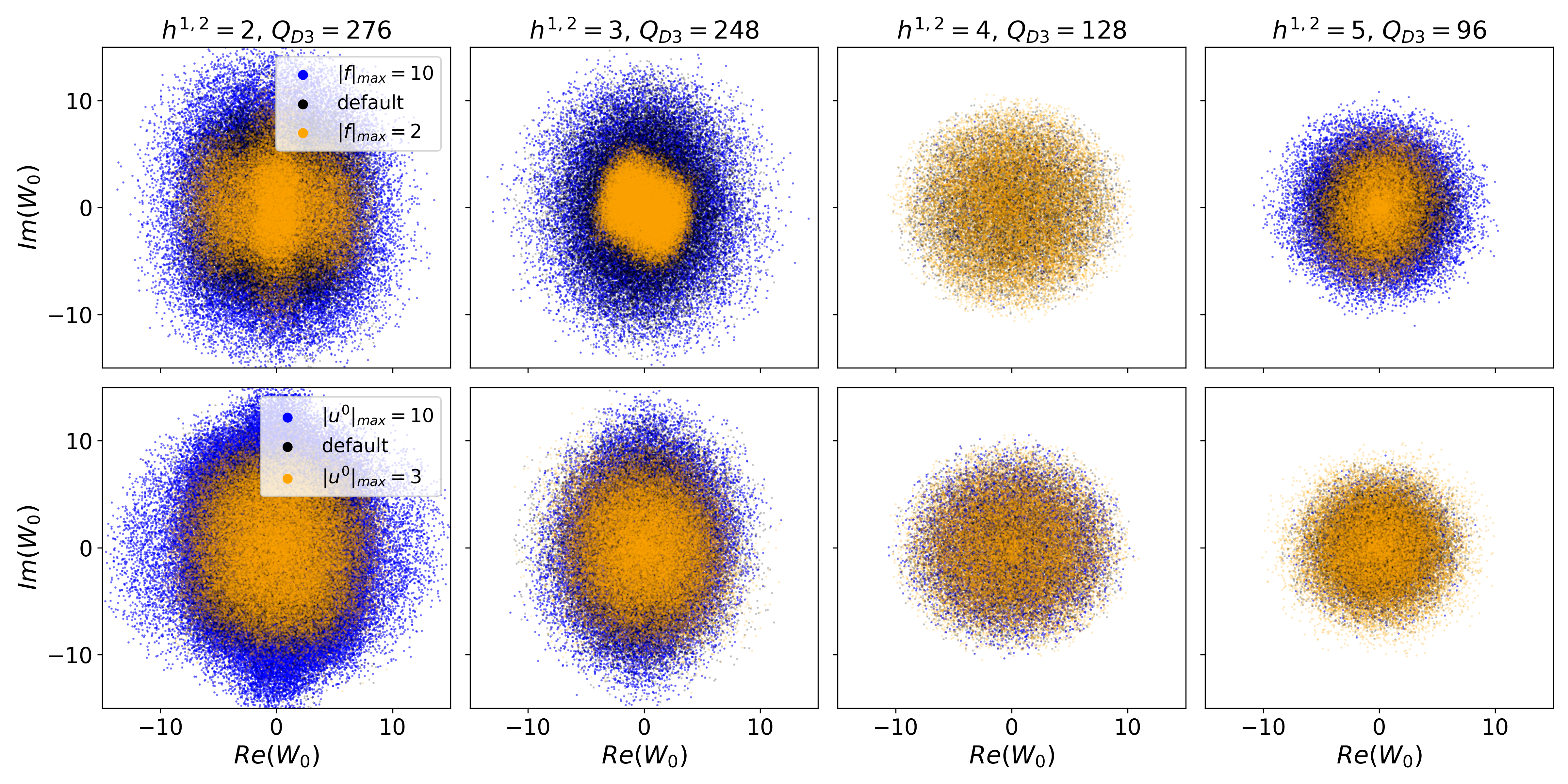}
    \caption{Dependence of the distribution on algorithmic hyperparameter choices. \emph{Top}: The top line shows the influence on our choice for the initial flux vector, i.e.~we varied the sampling region for the respective subset of the flux vector. The different subplots show effect for different number of moduli. The default value is $5.$ \emph{Bottom}: The bottom row compares different regions for the initial values for the moduli fields. The default value is $5.$ In both cases we see a smaller effect for larger number of moduli.}
    \label{fig:depndence-hyperparameters}
\end{figure}

Having analysed the functional behaviour of $W_0,$ it is important to understand its dependence on our algorithmic choices described in Sec.~\ref{sec:geometries}. At this point, the exact influence of the sampling choices in Eq.~\eqref{eq:samplingrange} on our numerical solutions is non-linear, though our results seem to be largely independent of the chosen sampling method by comparing the $W_0$ distributions obtained from the various sampling procedures discussed in \cite{Dubey:2023dvu}. Next, we performed some empirical checks to get a rough idea of the resilience of our results against variations of hyperparameters (see Fig.~\ref{fig:depndence-hyperparameters}). In particular, we change the range for the sampled flux vectors as well as the initial guesses for the moduli VEVs. Overall, we only observe mild effects. Nevertheless, our results are subject to these algorithmic assumptions (see~\cite{Krippendorf:2022gcl} for examples of such effects), but at this stage we are not aware of particular biases which are introduced by our algorithm. A more detailed analysis of algorithmic effects (e.g.~to examine the sub-leading features seen in the distribution) are beyond the scope of this paper.

\section{Conclusions}\label{sec:conclusions}

Our numerical methods allow a first glance at largely uncharted classes of string theory solutions. Studying these ensembles of flux vacua enables us to obtain meaningful insights into the distribution of $|W_0|$. Here, we were able to find a hint for universal features across geometries as the distributions exhibit rather similar properties. Among others, this includes a characteristic scaling of the width of our distributions with the D3-charge contribution $N_{\rm flux}$ from fluxes. Further, we observed a close similarity to a Gaussian distribution for $W_0$, which, to our knowledge, explicitly verifies for the first time the behaviour discussed in~\cite{Douglas:2003um,Ashok:2003gk,Denef:2004ze,Denef:2008wq}. This in turn also sheds a different perspective on finding solutions with small $|W_0|$ as it is equivalent to sampling small absolute values in a two-dimensional Gaussian.

Without more efficient sampling techniques, hierarchical suppression will require exponentially more samples. Learning strategies to efficiently generate such samples promise an exciting angle to study these particular regions of string theory solutions (see~\cite{Cole:2019enn,Krippendorf:2021uxu,Cole:2021nnt} for successfully learned strategies in simple string theory settings). This approach seems to be timely to complement human strategies to identify such special solutions~\cite{Demirtas:2019sip} at the tails of distributions within the landscape. For instance, it will be interesting to see how our numerical approach can be used to explicitly search for solutions with small tadpole at $h^{1,2}\gg 1$ -- often referred to as the tadpole conjecture~\cite{Bena:2020xrh} (see~\cite{Bena:2021wyr,Tsagkaris:2022apo} for numerical work in this direction).

It is clear from our work that a Gaussian distribution at the level of our samples seems to be a first approximation and that there are additional effects in our data which need to be accounted for. At this stage we were not able to attribute these additional features to algorithmic biases which therefore deserve further attention in the future. Some of us previously discussed potential sampling biases arising from the sampling method of initial points~\cite{Dubey:2023dvu} and of the algorithm itself~\cite{Krippendorf:2022gcl}. Ultimately, quantifying these biases is a prerequisite for a proper understanding of the distributions arising in the string landscape.

We note that an analysis of other phenomenologically relevant variables along the lines of our analysis of the distribution of $W_0$ seems straight-forward, but at this stage we do not see a pressing physics case for such an analysis. Arguably this situation is changed when Kähler moduli stabilisation is included (e.g.~\cite{Cicoli:2013swa}) as this sets many scales for the underlying particle physics model.

Another important line of future investigation is the comparison with random matrix potentials to understand the nature of the underlying string theory ensemble (see~\cite{Marsh:2011aa, Bachlechner:2012at} for work on random matrix potentials and random potentials inspired by flux vacua at LCS \cite{Brodie:2015kza,Marsh:2015zoa} for spectral properties of the Hessian).

Clearly the analysis presented in this note is by no means exhaustive and the hints for universal behaviour deserve further study. We hope to report on progress along these lines at a later stage.

\subsection*{Acknowledgments}
We thank Michele Cicoli, Nico Hamaus, Liam McAllister, Erik Plauschinn, Fernando Quevedo and Jochen Weller for useful discussions. We especially thank Andres Rios-Tascon for providing the code to compute GV and GW invariants. AS thanks the Ludwig Maximilian University of Munich and ICISE in Quy Nhon, Vietnam, for hospitality where parts of this work have been completed. The research of AS is supported by NSF grant PHY-2014071. 

\bibliographystyle{utphys}
\bibliography{bibliography}

\end{document}